\documentclass[intlimits,twoside,a4paper]{article}

\usepackage{amsmath,amssymb}
\usepackage{graphicx}
\usepackage{wrapfig}

\usepackage[T2A]{fontenc}
\usepackage[cp1251]{inputenc}

\usepackage[eqsecnum]{cmpj2}

\issue{2014}{17}{4}{43301}
\doinumber{10.5488/CMP.17.43301}

\title[The electronic properties of doped SWCNT and CNT sensor]%
{The electronic properties of doped single walled carbon nanotubes and carbon nanotube sensors}
\author[E. Tetik]{E.~Tetik}
\address{
Mustafa Kemal University, Computer Science Research and Application Center, Serinyol Campus, \\ 31100 Hatay, Turkey
}

\date{Received March 23, 2014, in final form September 29, 2014}
\authorcopyright{E. Tetik, 2014}

\begin{document}

\maketitle

\begin{abstract}
We present ab initio calculations on the band structure and density of states of single wall semiconducting carbon nanotubes with high degrees (up to 25\%) of B, Si and N substitution. The doping process consists of two phases: different carbon nanotubes (CNTs) for a constant doping rate and different doping rates for the zigzag (8, 0) carbon nanotube. We analyze the doping dependence of nanotubes on the doping rate and the nanotube type. Using these results, we select the zigzag (8, 0) carbon nanotube for toxic gas sensor calculation and obtain the total and partial densities of states for CNT (8, 0). We have demonstrated that the CNT (8, 0) can be used as toxic gas sensors for CO and NO molecules, and it can partially detect Cl$_2$ toxic molecules but cannot detect H$_2$S. To overcome these restrictions, we created the B and N doped CNT (8, 0) and obtained the total and partial density of states for these structures. We also showed that B and N doped CNT (8, 0) can be used as toxic gas sensors for such molecules as CO, NO, Cl$_2$ and H$_2$S.
\keywords ab initio calculations, carbon nanotubes structure, gas sensors, doping and substitution effects
\pacs 31.15.A-, 61.48.De, 07.07.Df, 74.62.Dh
\end{abstract}

\section{Introduction}

The physics of carbon nanotubes (CNTs) has evolved into a research field since their discovery \cite{1,2}. It is well known that carbon nanotubes could have either metallic or semiconducting properties depending on their diameters and chiralities \cite{3,4,5,6,7,8,9}. In this regard, carbon nanotubes have been recommended for several potential applications such as molecular sensors, hydrogen storage and nano-electronic devices. In view of incorporating CNTs into real operational nano-devices (diodes, transistors), CNT-based intra-molecular junctions have been studied for a long time \cite{10,11,12,13}. The most important point in these applications of carbon find efficient ways to control and regulate their structural and electronic properties. Foreign atom doping can be used as a feasible way to control the electrical and electronic properties of nanotubes.

Starting from the discovery of nanotubes, many papers have been published reporting the effects of their
electronic properties by direct adsorption of atoms on CNTs \cite{14,15,16}. Zhang et al.
systematically studied the isomers of BN doped on (5, 5) armchair carbon nanotubes and found
that the doping increases the redox and electron excitation properties \cite{17}. Krainara et al.
investigated the quantum chemical calculations to study the electronic properties of BN-doped carbon
nano-materials grafted with $n$-nucleophiles. They used used the {PBE} density functional theory (DFT)
method with the double-$con$ with polarization basis set and the {RI} approximation and examined that
the charge redistribution shows charge transfer when the BN-doped {SWCNT} is grafted with $m$-nitroaniline
and pyridine \cite{18}. Recently the researchers revealed many new research areas on the CNTs.
One of them contained the studies which were performed using doped CNTs to detect the presence of some toxic chemical gases.
CNTs have the potential to be developed as a new gas sensing material due to their inherent properties such as their small
size, great strength, high electrical and thermal conductivity, high surface-to-volume ratio, and hollow structure of nanomaterials.
These advances have led to the design of a new breed of sensor devices. In addition, the characteristics of CNTs have been
investigated for gas molecules adsorption, such as boron doped and boron nitride CNT.

In this work, we have reported a detailed study on the doping effect on the CNTs using ab initio DFT. We have investigated the calculations in three steps. In the first step, we changed the carbon nanotube chirality for the constant doping rate and found out how electronic properties of CNTs affect different types of constant doping nanotubes. In the second step, we changed the doping rate for a specific nanotube and examined the effect of doping rate on the nanotube. We used the atoms which are commonly selected such as B, Si and N for the doping process. Thus, we performed a comprehensive analysis examining the effects of doping on carbon nanotubes. In the final step, we identified a suitable nanotube for doping and obtained a toxic nano-gas-sensor that can detect the presence of H$_2$S, Cl$_2$, CO and NO molecules. Then, we examined the adsorption behaviors of the proposed pure and doped CNTs, in this regard, calculated the adsorption energy and binding distance using the total and partial densities of states for the CNT structures considered. We have demonstrated that these gas sensors are capable not only of detecting the presence of these toxic gases but also the sensitivity of these sensors can be controlled by the doping level of impurity atoms in a CNT.

\section{The theory and computational method}

In our work we used two programs which are the Tubegen code \cite{19} and SIESTA ab initio package \cite{16,21}.
We relaxed the nanotubes by using the \texttt{radius\_conv}, \texttt{error\_conv} and \texttt{gamma\_conv} which are parameters
in the Tubegen code and obtained the atomic position and the lattice parameters which were used in the SIESTA input file.

Total energy and electronic structure calculations are performed via first principles density functional theory, as implemented in the SIESTA. For the solution method we used the diagon which is a parameter in the SIESTA. For the exchange and correlation terms, the local density approximation (LDA) was used \cite{22} through the Ceperley and Alder functional \cite{23} as parameterized by Perdew and Zunger \cite{24}. The interactions between electrons and core ions are simulated with separable Troullier-Martins \cite{25} norm-conserving pseudo-potentials. We generated atomic pseudopotentials separately for atoms, C, B, N, Si and O by using the $2s^22p^2$, $2s^22p^1$, $2s^22p^3$, $3s^23p^2$ and $2s^22p^4$ configurations, respectively. For carbon atoms, $1s$ state is the core state while the $2s$ and $2p$ states form the valence states. The valence electrons were described by localized pseudo-atomic orbitals with a double-$con$ singly polarized (DZP) basis set \cite{26}. Basis sets of this size have been shown to yield structures and total energies in good agreement with those of standard plane-wave calculations \cite{27}. Real-space integration was performed on a regular grid corresponding to a plane-wave cutoff around 300~Ry, for which the structural relaxations and the electronic energies are fully converged. We used 25 $k$ points for the total energy calculations of pure and doped CNTs. We relaxed the isolated CNTs until the stress tensors were below 0.04~eV/A$_{con3}$ and calculated the theoretical lattice constant. In the band structure calculations we used 161 band $k$ vectors between the $con$ and $A$ high-symmetry points.

\section{Results and discussion}

\begin{figure}[!t]
\centerline{
\includegraphics[width=0.5\textwidth]{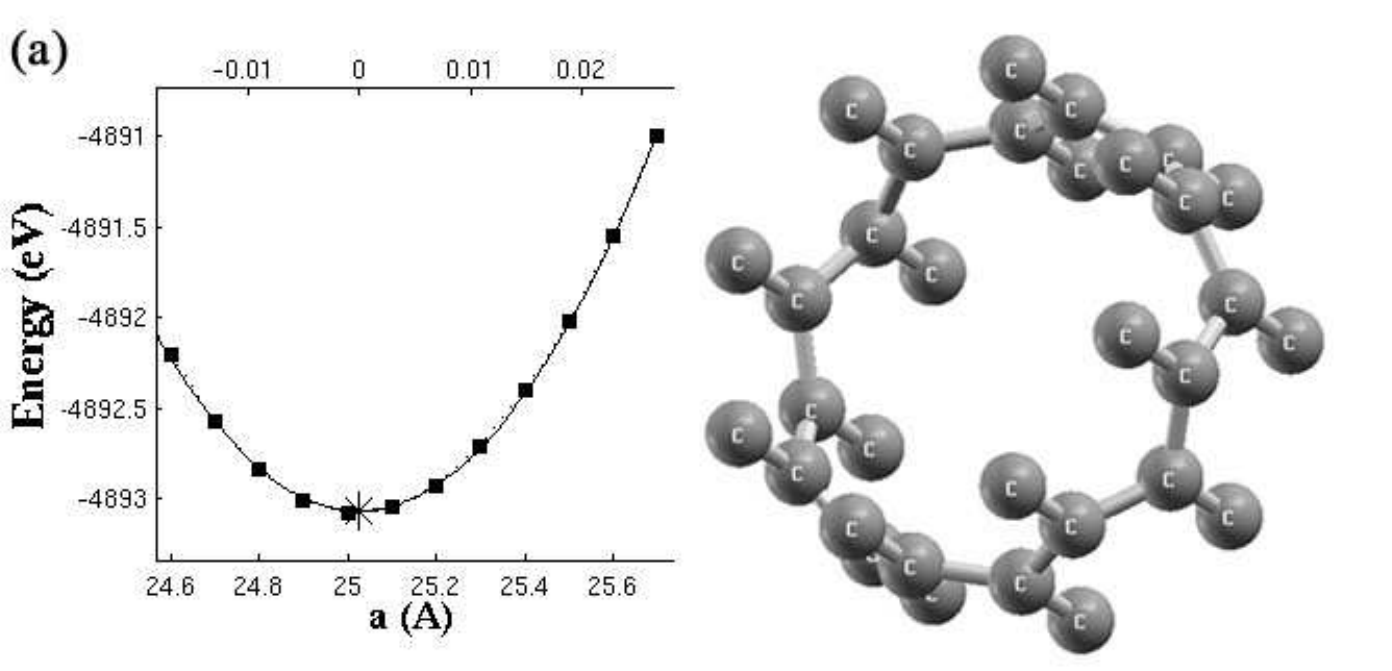}
\includegraphics[width=0.5\textwidth]{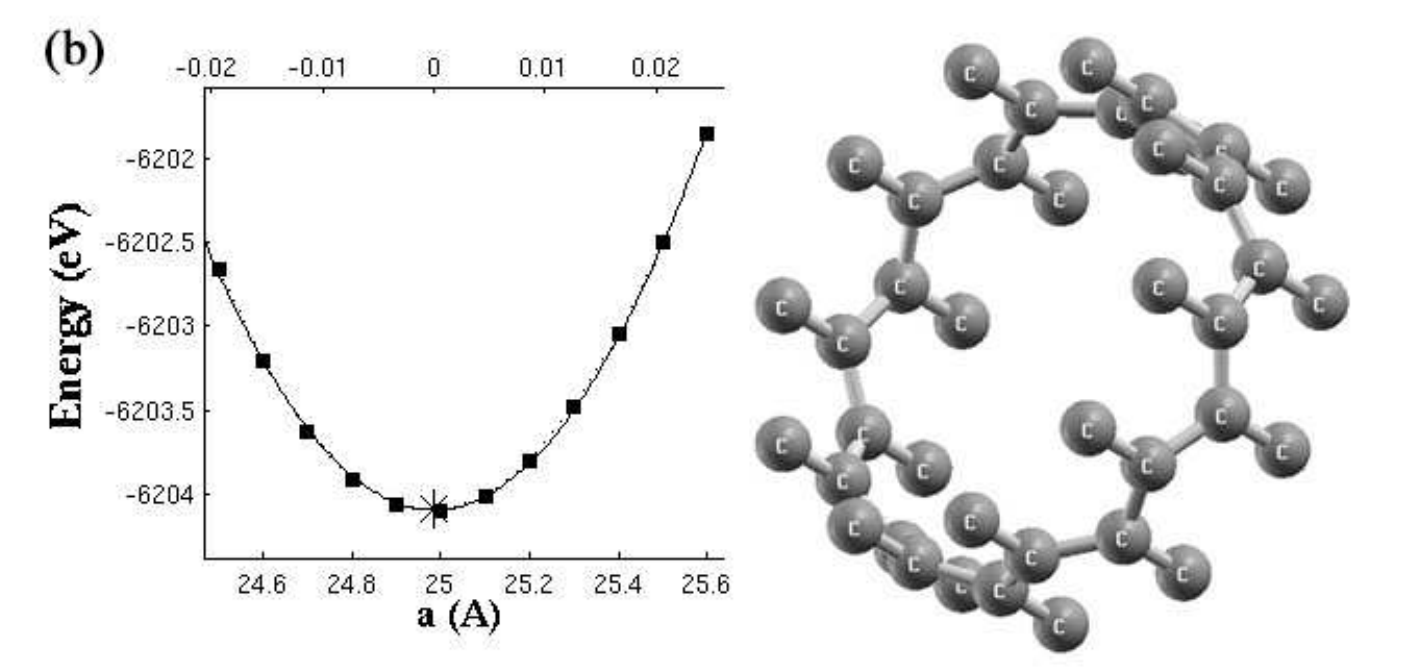}
}
\centerline{
\includegraphics[width=0.5\textwidth]{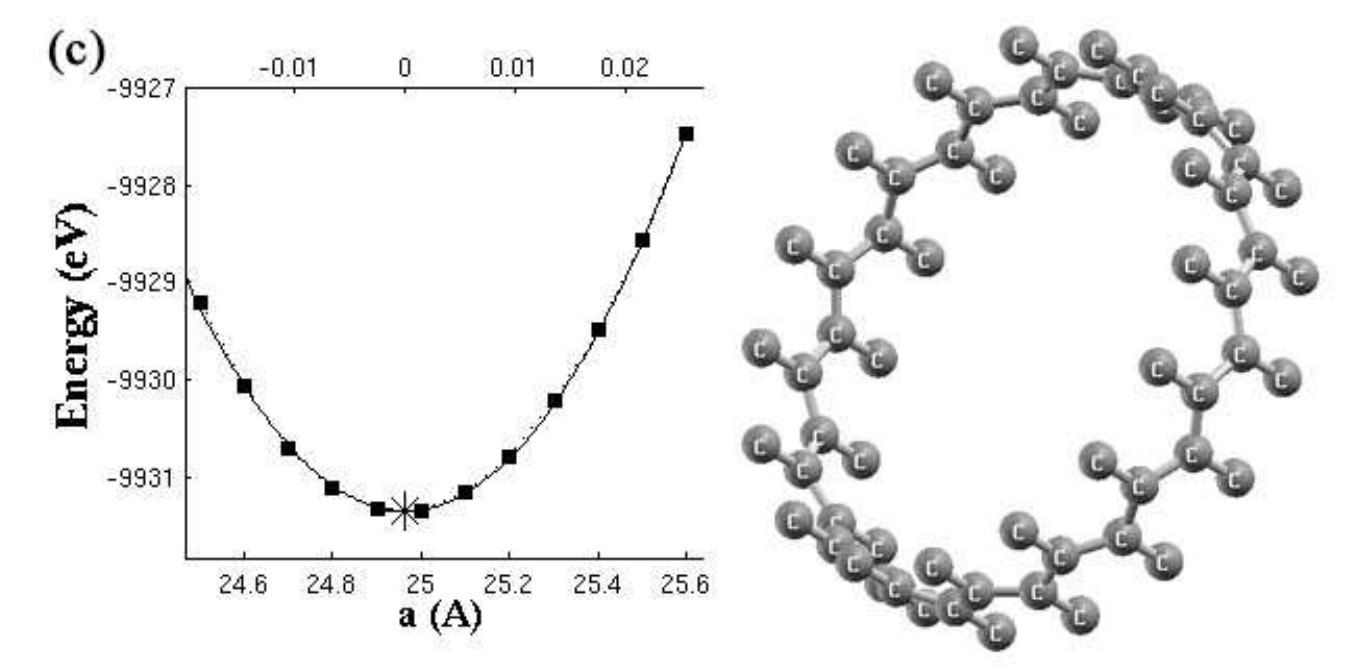}
}
\caption{The graphics of equilibrium lattice constant  and 3D diagram for CNT (8, 0) (a), CNT (10, 0) (b) and CNT (16, 0) (c). } \label{fig1}
\end{figure}

We can say that all physical properties are related to the total energy. For example,
the equilibrium lattice constant of a structure is the lattice constant that
minimizes the total energy. If the total energy is calculated, any physical property
related to the total energy can be determined. In this regard, we have relaxed the
zigzag (8, 0), (10, 0) and (16, 0) and obtained their equilibrium lattice parameters
which have been computed minimizing the total energy of crystals calculated for
different values of lattice constant. The calculation results of the zigzag (8, 0), (10, 0) and (16, 0)
CNTs are shown in figure~\ref{fig1}, and the equilibrium lattice parameters for these CNTs are found to
be $a_{\mathrm{CNT}(8,0)} = 25.0227$~\AA{}, $a_{\mathrm{CNT}(10,0)} = 24.9851$~\AA{} and
$a_{\mathrm{CNT}(16,0)}= 24.9644$~\AA{}, respectively.

Then, we created the B, Si and N doped CNTs for constant and different doping rate. We obtained the electronic properties of doped CNTs that were classified according to the doping rate. We discussed the band structure and density of state of pure and doped CNTs. The properties of doped CNTs changed according to the doped atoms and the doping rate. Our ultimate goal here is to predict the exact behavior of nanotubes depending on the doping rate and the doped atom. According to these results, the most appropriate nanotube can be determined for other studies. Finally, we studied the interaction of CO, NO, H$_2$S and CI$_2$ molecules with CNT (8, 0). We found that the partial densities of states of the pure and doped zigzag CNT (8, 0) altered considerably after detecting the CO, NO, H$_2$S and CI$_2$ molecules.

\subsection{Different carbon nanotubes for a constant doping rate}

\begin{figure}[!b]
\centerline{\includegraphics[width=0.98\textwidth]{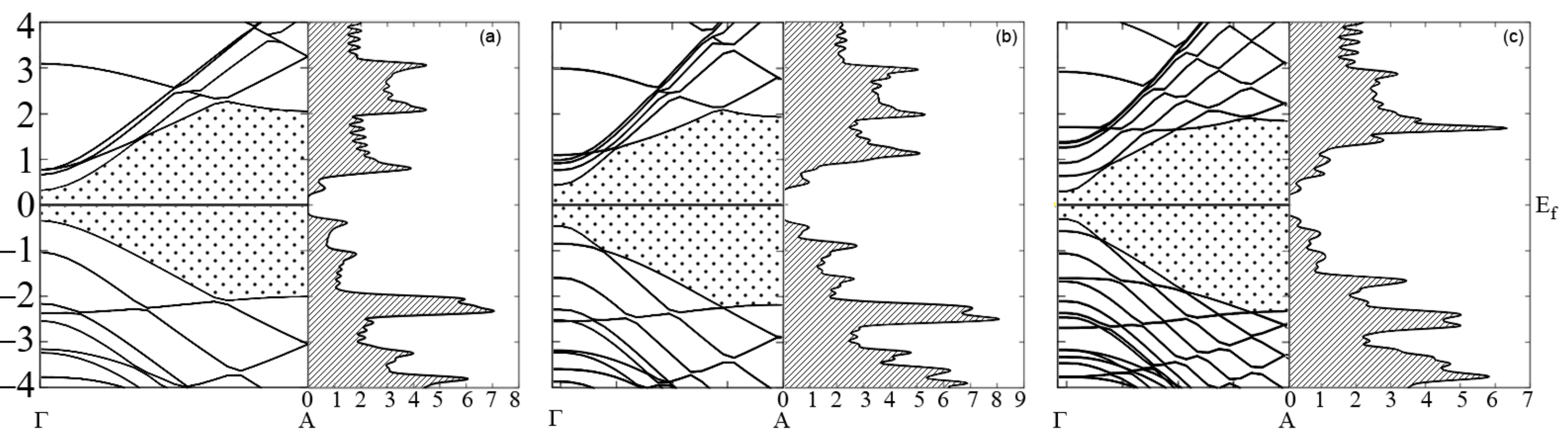}}
\caption{Band structure and density of state of zigzag (a) (8, 0), (b) (10, 0) and (c) (16, 0) nanotubes.} \label{fig2}
\end{figure}

In figure~\ref{fig2} we show the band structure and density of state (DOS) for pure zigzag (8, 0), (10, 0) and (16, 0)
nanotubes that include 32, 40 and 64 carbon atoms, respectively. The Fermi level is set to zero in figure~\ref{fig2}
and in all other graphics of band structure and density of state. We can see that all zigzag nanotubes exhibit
semiconductor properties. Their semiconducting energy gaps are $E_\textrm{g}= 0.6643$, $0.8927$ and $0.5914$~eV,
respectively (table~\ref{tbl-smp1}).  Pure zigzag (10, 0) nanotube has the largest band gap.

\begin{figure}[!t]
\centerline{\includegraphics[width=0.98\textwidth]{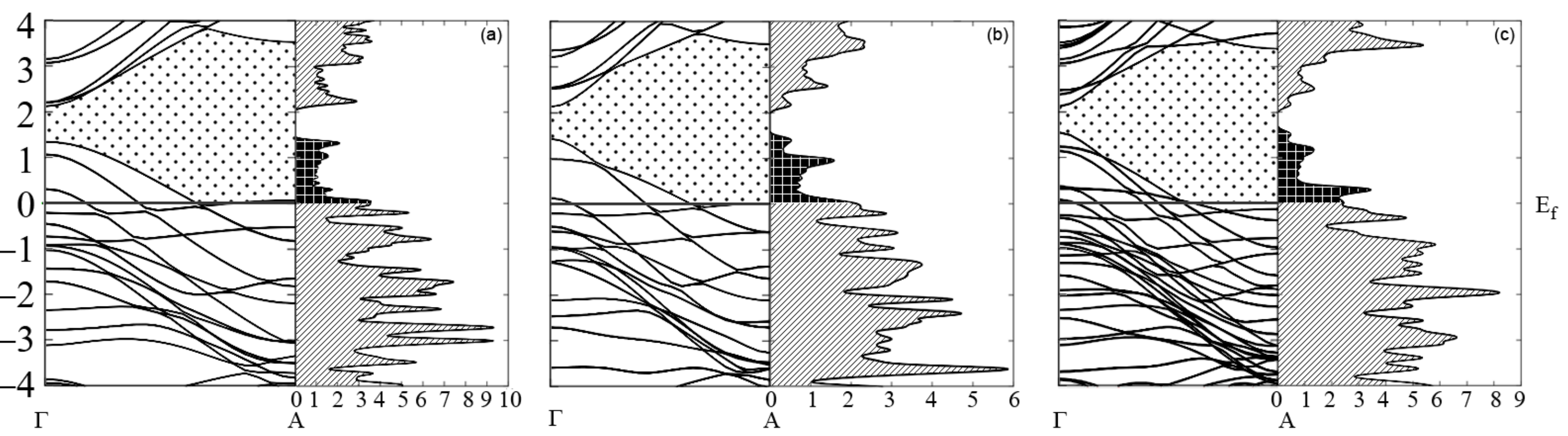}}
\caption{Band structure and density of state of 12.5\% B doped zigzag (a) (8, 0), (b) (10, 0) and (c) (16, 0) nanotubes.} \label{fig3}
\end{figure}

Figure~\ref{fig3} contains the electronic dispersion and DOS for B doped zigzag (8, 0), (10, 0) and (16, 0)
nanotubes that include 4, 5 and 8 B atoms, respectively (table~\ref{tbl-smp1}).
Doping rate of these nanotubes is 12.5\%. We have used a high doping rate because we can see more clearly
the effects of doping. If we examine the DOS for three different semiconducting nanotubes,
we show that the peaks are near the Fermi level. Those peaks display the 1-D Van Hove Singularities
(VHSs) pattern. For undoped nanotubes, the DOS is symmetric around the band gap for the 1st and 2nd VHSs.
However, beyond this small energy regime around the band gap, asymmetries arise due to the mixing of $p$
and $s$ orbitals.  B doping leads to lowering of the Fermi level into the valence band of the undoped tube.
Above the highest occupied bands of an undoped tube, new bands are formed. These bands correspond to the
formation of an acceptor level in semiconductors having very low dopant concentration. The shift of the
Fermi level for the zigzag \mbox{(8, 0)}, (10, 0) and (16, 0) nanotube is around 1.36076~eV, 1.43003~eV and 1.55763~eV,
respectively (table~\ref{tbl-smp2}). Due to Fermi shift, the number of free electrons in a conduction band of B
doped nanotubes increases. Therefore, conductivity of these nanotubes increases as well.

\begin{table}[!h]
\caption{The band gap (eV) of pure and doped (B, Si and N) zigzag (8, 0), (10, 0) and (16, 0) nanotube.}
\label{tbl-smp1}
\vspace{1ex}
{\small{
\begin{center}
\renewcommand{\arraystretch}{0}
\begin{tabular}{|c||c|c|c|c|c|c|}
\hline
Material&Pure (present)&Pure (reference)&B Doped&Si Doped&N Doped&CNT/Doped Atoms\strut\\
\hline
\rule{0pt}{2pt}&&&&&&\\
\hline
      CNT (8, 0)& --0.6643& --0.47 \cite{28}&  --0.7761&  --0.4929&  --0.8263&  32/4\strut\\
\cline{1-7}
      CNT (10, 0)& --0.8927& --0,75 \cite{29}, --0.98 \cite{29}&  --0.6872&  --1.0838&  --0.7702&  40/5\strut\\
\hline
      CNT (16, 0)& --0.5914& --0.62 \cite{30}&  --0.5634&  --1.0089& --0.6430& 64/8\strut\\
\hline
\end{tabular}
\renewcommand{\arraystretch}{1}
\end{center}}}
\end{table}

\begin{table}[!b]
\caption{The shift of the Fermi level for the B, Si and N doped zigzag (8, 0), (10, 0) and (16, 0) nanotube.}
\label{tbl-smp2}
\vspace{1ex}
{\small{
\begin{center}
\renewcommand{\arraystretch}{0}
\begin{tabular}{|c||c|c|c|}
\hline
Material&B Doped (A)&Si Doped (A)&N Doped (A)\strut\\
\hline
\rule{0pt}{2pt}&&&\\
\hline
      CNT (8, 0)& 1.36076& 0.0&  --0.95805\strut\\
\cline{1-4}
      CNT (10, 0)& 1.43003& 0.0&  --1.03256\strut\\
\hline
      CNT (16, 0)& 1.55763& 0.0&  --1.23198\strut\\
\hline
\end{tabular}
\renewcommand{\arraystretch}{1}
\end{center}}}
\end{table}

\begin{figure}[!t]
\centerline{\includegraphics[width=0.98\textwidth]{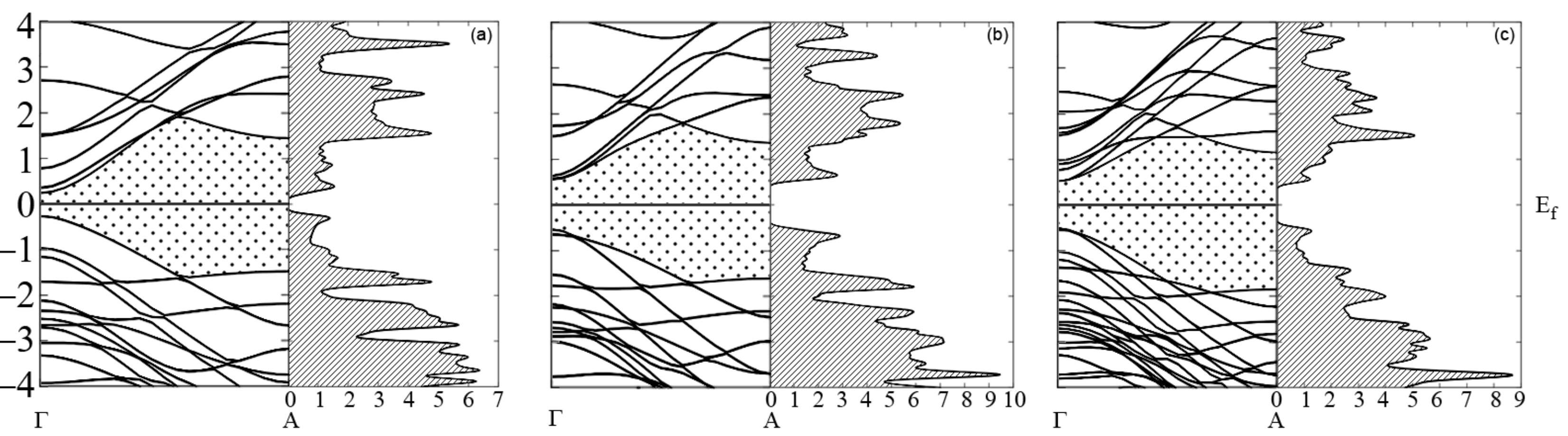}}
\caption{Band structure and density of state of 12.5\% Si doped zigzag (a) (8, 0), (b) (10, 0) and (c) (16, 0) nanotubes.} \label{fig4}
\end{figure}

\begin{figure}[!t]
\centerline{\includegraphics[width=0.98\textwidth]{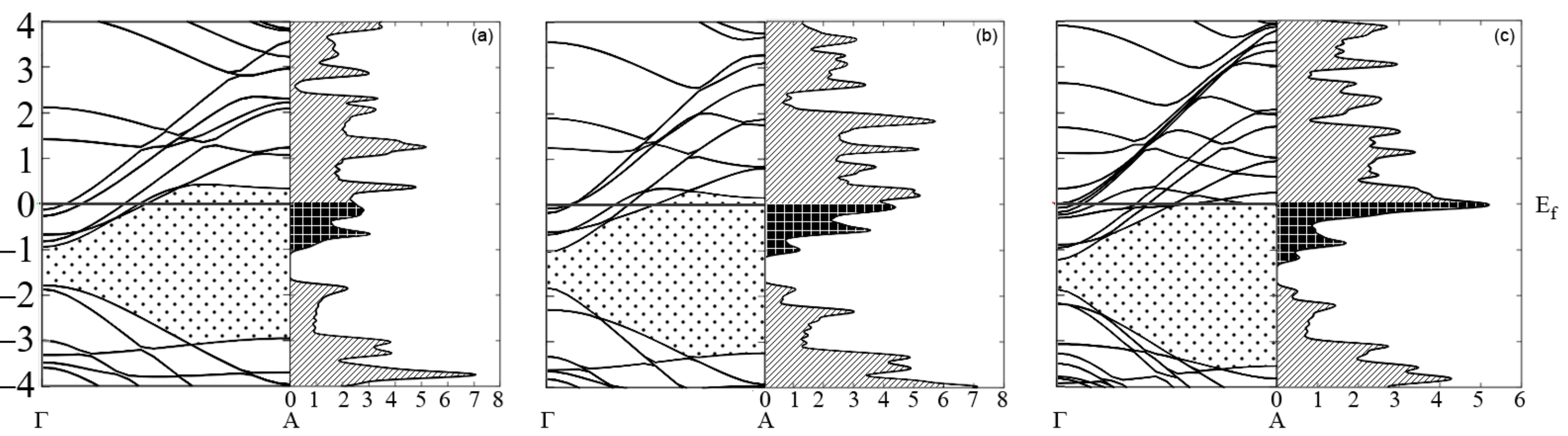}}
\caption{Band structure and density of state of 12.5\%  N doped zigzag (a) (8, 0), (b) (10, 0) and (c) (16, 0) nanotubes.} \label{fig5}
\end{figure}

To investigate the effects of doping on the electronic properties of nanotubes, we have performed similar calculations for zigzag (8, 0), (10, 0) and (16, 0) nanotube using Si and N atoms. We show the electronic band structure and DOS of Si and N doped nanotubes in figures~\ref{fig4} and \ref{fig5}, respectively.  Carbon and silicon are in the same periodic group and the electron configurations are very close to each other. Therefore, the Si doped nanotubes do not have a shift of the Fermi level. We show that the band gap of the Si doped nanotubes changed significantly (table~\ref{tbl-smp2}). The band gap of these nanotubes is $-0.4929$~eV, $-1.0838$~eV and $-1.0089$~eV and is fixed around 1~eV. There is a great change in the band gap of the N doped nanotubes. As we can see in figure~\ref{fig5}, the shift of the Fermi level for the N doped tubes is around $0.95805$~eV, $1.03256$~eV and $1.23198$~eV, respectively (table~\ref{tbl-smp2}), which is a clear indication that the number of free free electrons of N doped semiconducting nanotubes decreases. As a result, these nanotubes are a finite conductance and form N-type %


\begin{table}[!b]
\caption{The band gap of doped (B, Si and N) zigzag (8, 0) nanotube for Different Doping Rates.}
\label{tbl-smp3}
\vspace{1ex}
{\small{
\begin{center}
\renewcommand{\arraystretch}{0}
\begin{tabular}{|c|c|c|c|c|}
\hline
Material& 3.125\% (A) (1 atom)& 6.25\% (A) (2 atoms)& 12.5\% (A) (4 atoms)& 25\% (A) (8 atoms)\strut\\
\hline
\rule{0pt}{1pt}&&&&\\
\hline
      CNT-B (8, 0)& --0.2862& --0.3831&  --1.3984& --2.0600\strut\\
\cline{1-5}
      CNT-Si (8, 0)& --0.4529& --0.4529&  --0.4929& --0.9292\strut\\
\hline
      CNT-N (8, 0)& --0.3973& --0.2729&  --0.8263& --0.9521\strut\\
\hline
\end{tabular}
\renewcommand{\arraystretch}{1}
\end{center}}}
\end{table}

\subsection{Different doping rates for zigzag (8, 0) carbon nanotubes}

\begin{figure}[!t]
\centerline{\includegraphics[width=0.65\textwidth]{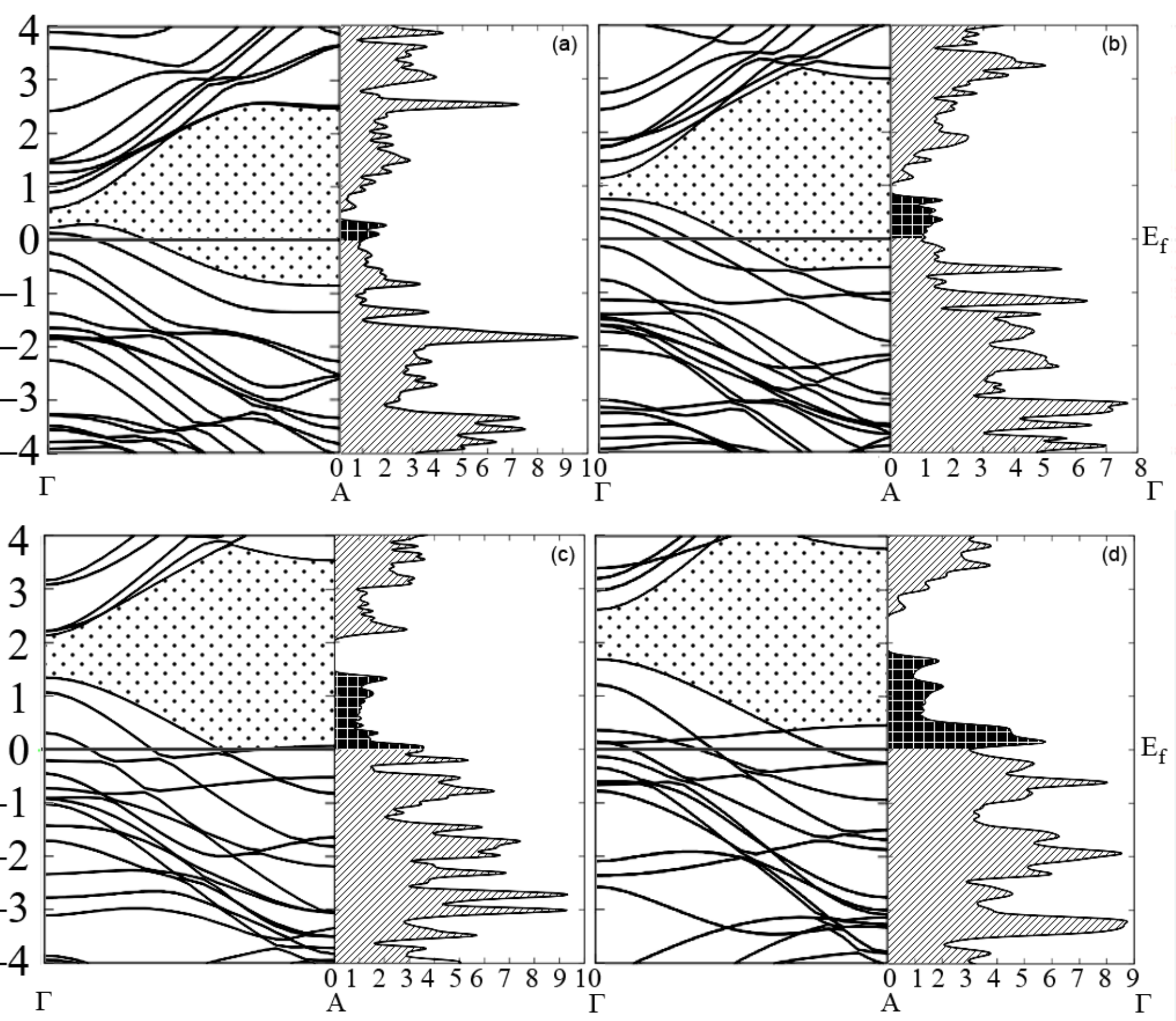}}
\caption{Band structure and density of state of (a) 3.125\%, (b) 6.25\%, (c) 12.5\% and (d) 25\% B doped zigzag (8, 0) nanotubes.} \label{fig6}
\end{figure}

\begin{figure}[!b]
\centerline{\includegraphics[width=0.65\textwidth]{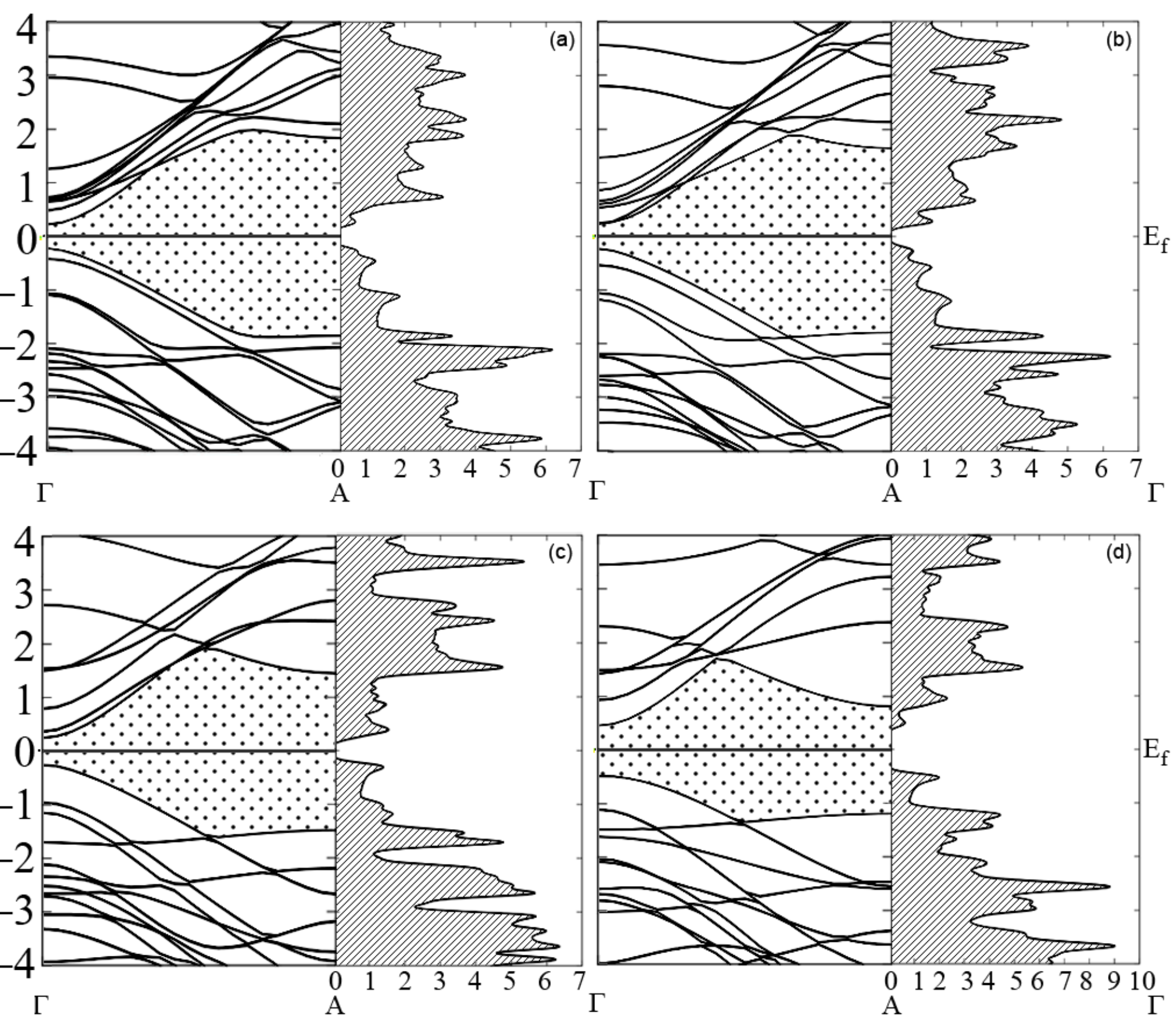}}
\caption{Band structure and density of state of (a) 3.125\%, (b) 6.25\%, (c) 12.5\% and (d) 25\% Si doped zigzag (8, 0) nanotubes.} \label{fig7}
\end{figure}

Secondly, we have studied the electronic band structure and DOS for a zigzag (8, 0) nanotube using different
doping rates (3.125\%, 6.25\%, 12.5\% and 25\%). We have calculated that the numbers of B atoms
corresponding to these doping rates are 1, 2, 4 and 8, respectively. The results for the doped zigzag (8, 0)
nanotubes which include 1, 2, 4 and 8 B atoms (according to the doping rate) are presented in table~\ref{tbl-smp3}.
Moreover, we show the band structure and DOS of the B doped zigzag (8, 0) nanotubes in figure~\ref{fig6}
according to the doping rate. If we analyze these results, the number of free electrons of nanotubes
increases when the doping rate is increased. On the other hand, the Fermi level shifts downward
into the conduction band of the doped semiconducting zigzag (8, 0) nanotubes according to the doping rate.
Due to the shift of the Fermi level the boron levels hybridize with the carbon levels forming highly
dispersive acceptor-like bands and consequently lending a metallic property to the nanotubes.
In addition to this study, we use Si and N atoms which are commonly used in the doping process
for nanotubes since we can show the effect on the doping rate of different atoms. The calculation
results regarding the band gap of CNT-Si and CNT-N are shown in table~\ref{tbl-smp1}. Then, we
obtain the changes in the band structure and DOS of Si doped nanotube for different doping rates
in figure~\ref{fig7}. We show that the band gap of the Si doped (8, 0) nanotubes increases according
to the doping rate. When the doping rate becomes closer to 25\%, the band gap greatly increases.
Although the doping rate is changed, it is seen that the Fermi level of the Si doped (8, 0)
nanotubes does not shift. Since C and Si atoms belong to the same group, this effect has arisen.


\begin{figure}[!h]
\centerline{\includegraphics[width=0.65\textwidth]{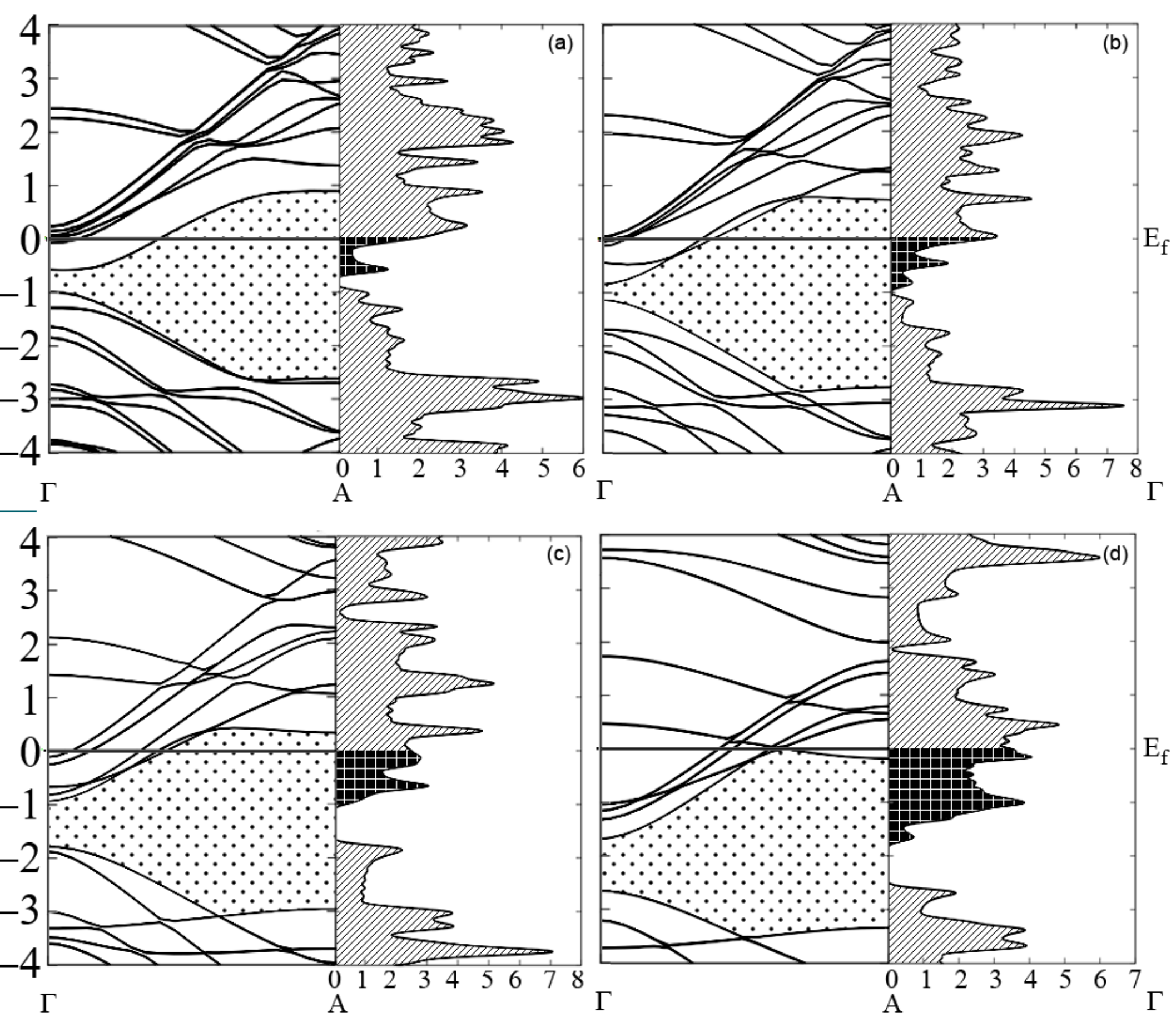}}
\caption{Band structure and density of state of (a) 3.125\%, (b) 6.25\%, (c) 12.5\% and (d) 25\% N doped zigzag (8, 0) nanotubes.} \label{fig8}
\end{figure}

The band structure and DOS of N doped nanotubes for different doping rates is evaluated and graphical results are presented in figure~\ref{fig8}. In the N doped nanotubes, the Fermi level shifts upward into the valence band according to the doping rate. When the doping rate increases, the number of free electrons of these nanotubes decreases and, as a result, the conductance becomes weak. We can say that the shift of the Fermi level strongly depends on the geometry of the structure and the dopant concentration. The shifts of the Fermi level for all nanotubes are presented in table~\ref{tbl-smp4}. As a result, the group of the doped atom affects the position of the Fermi level. We have compared the calculation results with the SWCNT upon boron substitution research by G.~Fuentes et al. \cite{31}. G.~Fuentes examined the (16, 0) nanotube for different doping rates (0\%, 6.25\%, 12.5\% and 25\%) in the theoretical part of the research. According to the calculation results, the stronger the doping, the bigger the shift of the Fermi level is. For B doping rate between 0 and 25\%, the shift of the Fermi level reaches 2.2~eV. In this study, the level shifts for (8, 0) and (16, 0), CNT reaches 1.71~eV (25\%) and  1.55~eV (12.5\%), respectively. We show that both studies indicate similar results according to the shift level.

\begin{table}[!t]
\caption{The shift of the Fermi level of a doped (B, Si and N) zigzag (8, 0) nanotube for Different Doping Rates.}
\label{tbl-smp4}
\vspace{1ex}
{\small{
\begin{center}
\renewcommand{\arraystretch}{0}
\begin{tabular}{|c|c|c|c|c|}
\hline
Material& 3.125\% (A) (1 atom)& 6.25\% (A) (2 atoms)& 12.5\% (A) (4 atoms)& 25\% (A) (8 atoms)\strut\\
\hline
\rule{0pt}{1pt}&&&&\\
\hline
      CNT-B (8, 0)& 0.30635& 0.74464&  1.35138& 1.71388\strut\\
\cline{1-5}
      CNT-Si (8, 0)& 0.0& 0.0& 0.0& 0.0\strut\\
\hline
      CNT-N (8, 0)& --0.58724& --0.85205& --0.95805& --1.69396\strut\\
\hline
\end{tabular}
\renewcommand{\arraystretch}{1}
\end{center}}}
\end{table}

\subsection{Zigzag (8, 0) carbon nanotube sensors}

\begin{figure}[!t]
\centerline{\includegraphics[width=0.98\textwidth]{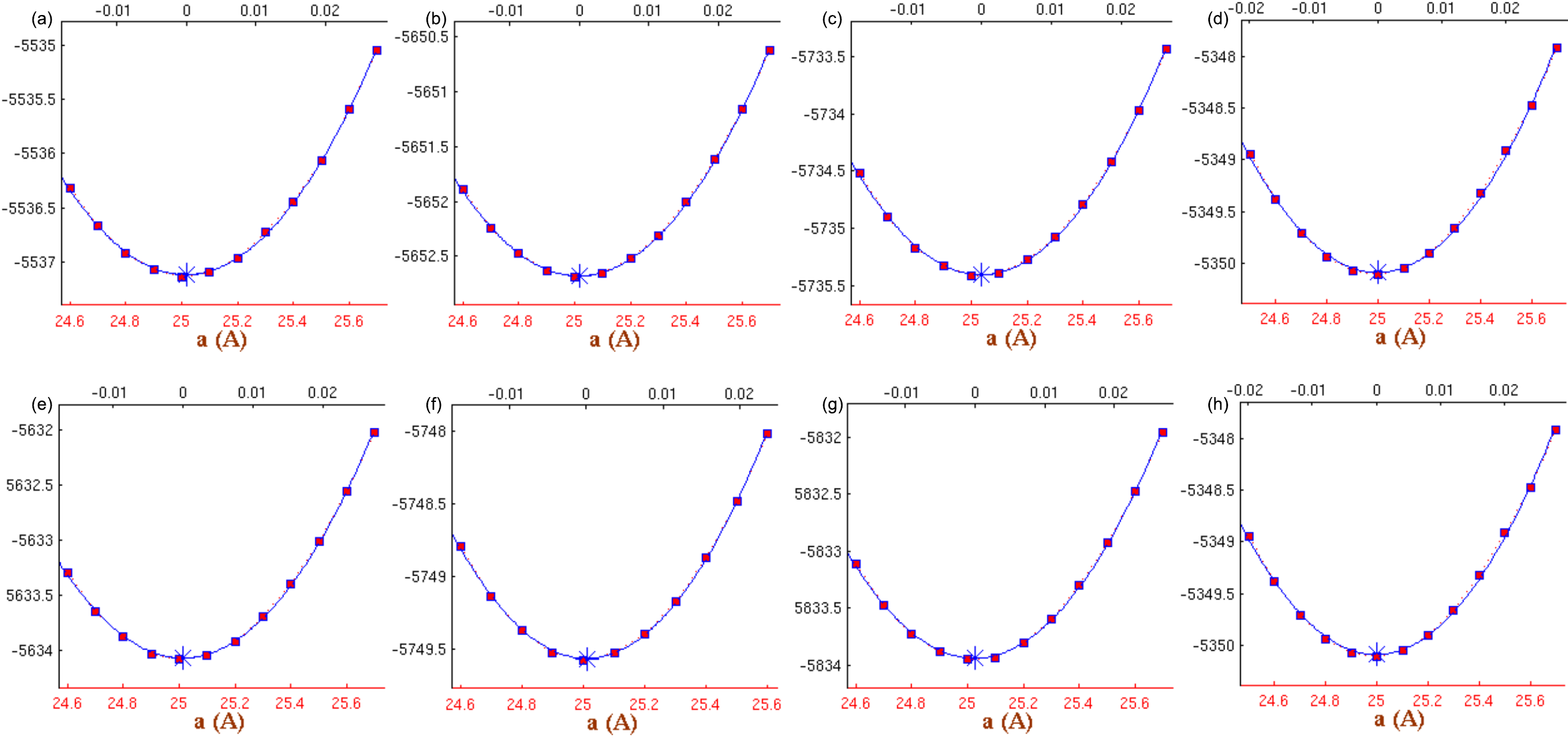}}
\caption{(Color online) The graphics of equilibrium lattice constant for pure CNT [CO (a), NO (b), Cl$_2$ (c) and H$_2$S (d)] and BN doped CNT [CO (e), NO (f), Cl$_2$ (g) and H$_2$S (h)].} \label{fig9}
\end{figure}

\begin{figure}[!b]
\centerline{\includegraphics[width=0.95\textwidth]{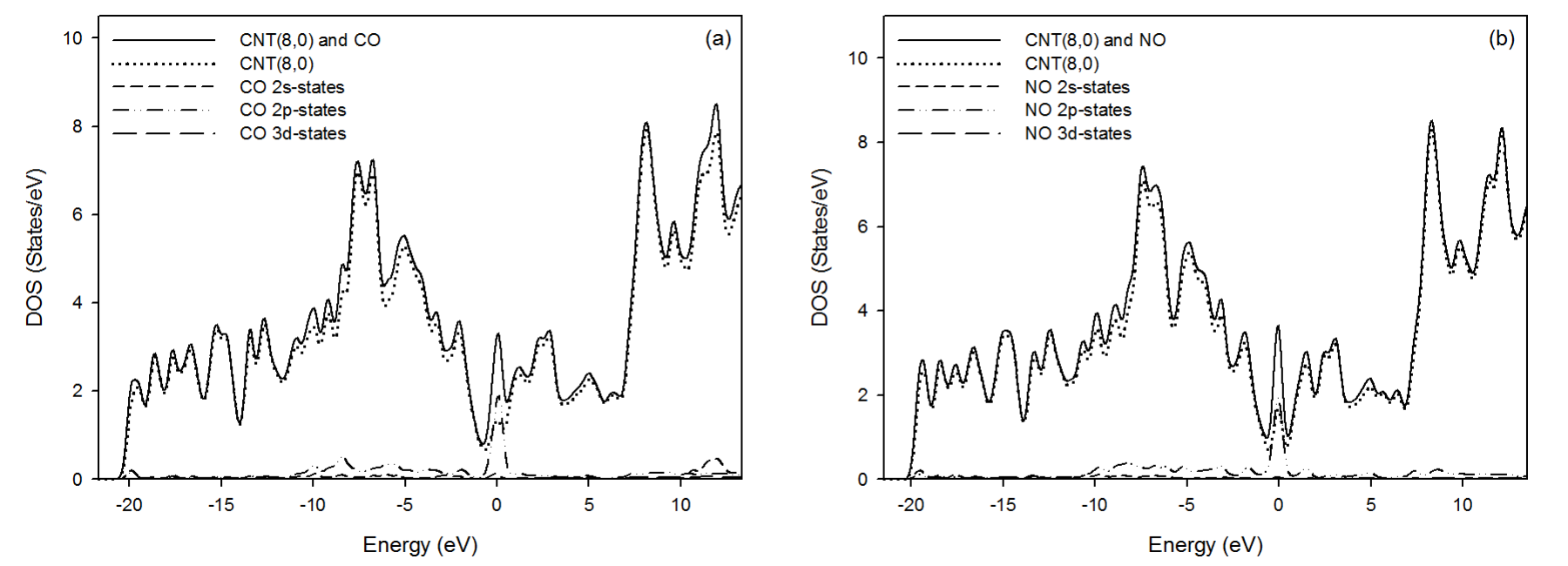}}
\caption{The total and projected density of states for adsorption of (a) CO and (b) NO on pure zigzag (8, 0) nanotube. The position of the Femi level is set to zero.} \label{fig10}
\end{figure}

In this section, it is proved that zigzag (8, 0) CNT is a suitable nanotube for gas sensor applications. We relaxed the zigzag (8, 0) and obtained their equilibrium lattice parameters for all toxic molecules (figure~\ref{fig9}). Since the changing properties of a nanotube after doping are quite stable, zigzag (8, 0) nanotube can be used as a gas sensor. In this respect, we investigate the interaction of CO, NO, Cl$_2$ and H$_2$S toxic molecules with a zigzag (8, 0) nanotube. The total and partial densities of states of toxic molecule and pure zigzag (8, 0) nanotube are illustrated in figures~\ref{fig9} and \ref{fig10}.

Partial densities of states consist of the $s$, $p$ and $d$ states of CNT (8, 0) and toxic molecule. In figure~\ref{fig9}, we show a remarkable peak in the Fermi level of CO and NO molecules. This peak is generated by the CO (a) and NO (b) toxic molecules. As we can see in figure~\ref{fig9}~(a), the lowest valence bands that occur between about $-20$ and $-11$~eV are dominated by CO (C and O total) $2s$ and $3d$ states while valence bands occurring between about $-10$ and $-2$~eV are dominated by CO $2p$ states. The highest occupied valence bands are essentially dominated by CO $2p$ states occurring between about $-2$ and $+2$~eV. Similar properties also apply to NO doped CNT (8, 0) nanotube in figure~\ref{fig9}~(b). For this nanotube, the highest occupied valence bands are dominated by NO $2p$ states occurring between about $-2.1$ and $+1.2$~eV. Since interaction nanotubes with CO and NO have a high peak in the Fermi level due to CO and NO $2p$ states, the conductivity of CNT (8, 0) nanotube considerably increases.

Secondly, we examine the interaction of Cl$_2$ and H$_2$S toxic molecules with a zigzag (8, 0) nanotube in figure~\ref{fig10}. In both studies, the lowest valence bands occur between about $-20$ and $-1$~eV which are dominated by $2s$, $2p$ and $3d$ states (see figure~\ref{fig10}) of Cl$_2$ and molecules. The states of Cl$_2$ molecule also contribute to the valence bands in the Femi level, but the values of densities of these states are very small compared to CO and NO $2p$ states. Therefore, we think that the possibility of applying the CNT (8, 0) nanotube as toxic gas sensor for Cl$_2$ is very low [figure~\ref{fig10}~(a)]. Figure~\ref{fig10}~(b) contains the interaction of H$_2$S toxic molecules with a zigzag (8, 0) nanotube. The states of H$_2$S molecule do not contribute to the valence bands in the Femi level. For this reason, CNT (8, 0) nanotube is not applicable for toxic gas sensors for H$_2$S.

\begin{figure}[!t]
\centerline{\includegraphics[width=0.95\textwidth]{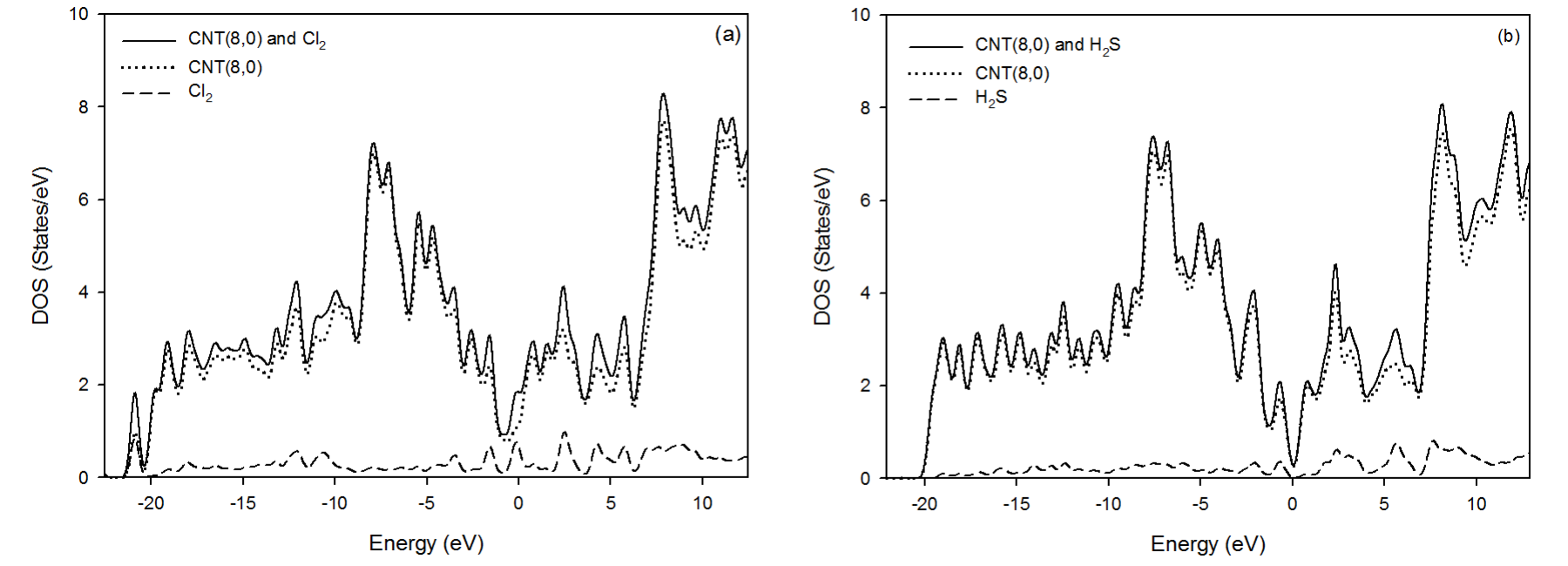}}
\caption{The total and projected density of states for adsorption of (a) Cl$_2$ and (b) H$_2$S on pure zigzag (8, 0) nanotube. The position of the Femi level is set to zero.} \label{fig11}
\end{figure}

\begin{figure}[!b]
\centerline{\includegraphics[width=0.95\textwidth]{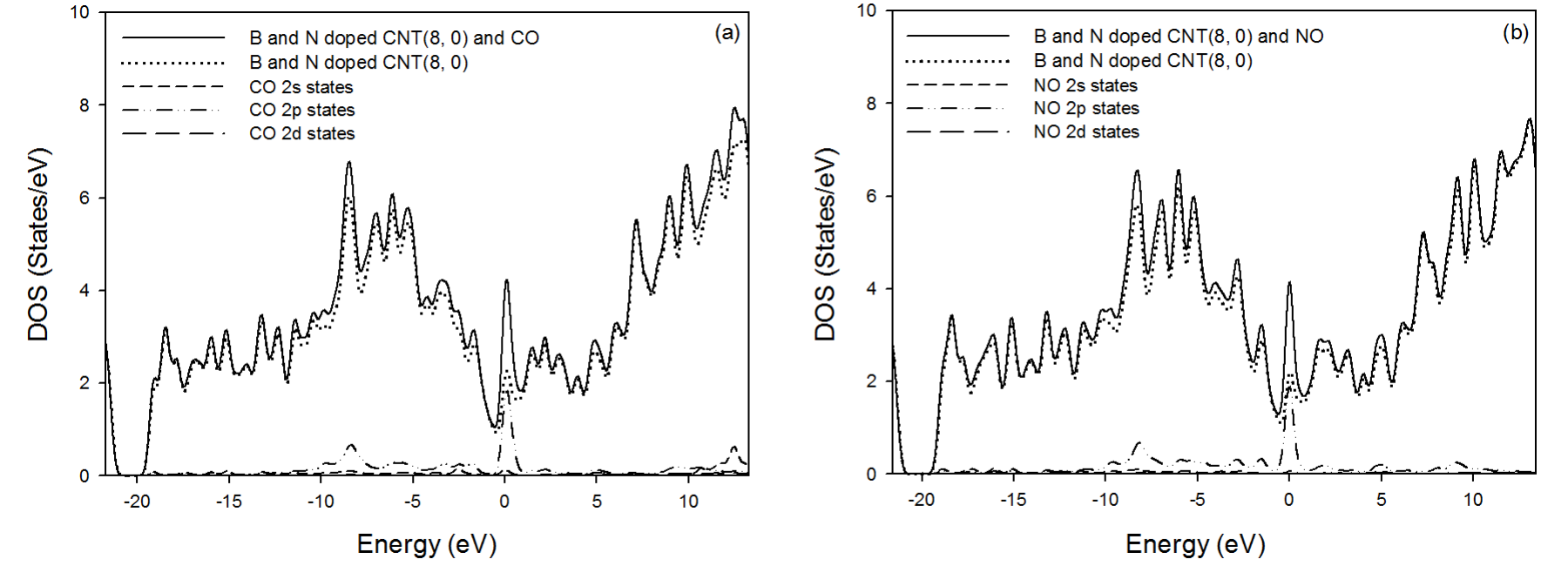}}
\caption{The total and projected density of states for adsorption of (a) CO and (b) NO on a doped zigzag (8, 0) nanotube. The position of the Femi level is set to zero.} \label{fig12}
\end{figure}

To investigate the effects of doping on toxic molecules, we optimize the B and N doped CNT (8, 0)
and realize a similar calculation for this structure. According to researches, undoped nanotubes
cannot detect all toxic gas molecules because some toxic molecules do not absorb on the surface
of a nanotube \cite{32}. To overcome these limitations of pure SWCNTs, diverse external or internal
process schemes can be used \cite{33}. We examine the interaction of CO, NO, Cl$_2$ and H$_2$S toxic
molecules with the B and N doped zigzag (8, 0) nanotube. The total and projected density of states
for these structures can be observed in figures~\ref{fig11} and \ref{fig12}, respectively. The results of calculation
are close to the results of a pure nanotube, but the peaks of a doped nanotube are more distinct and
higher than pure CNT (8, 0). Remarkable peaks in the Fermi level of Cl$_2$ and H$_2$S molecules are
shown in these figures. Similarly, conductivity of a doped CNT (8, 0) nanotube  considerably increases.
This is due to the interaction between the toxic molecules of a doped CNT. The $2p$ orbital of the nanotube
and toxic molecules create bonds which cause degenerate levels in a nanotube. In this case, the acceptor
levels of a nanotube occur, and it is concluded that doped nanotubes can be used as effective toxic gas
sensors for CO, NO, Cl$_2$ and H$_2$S (see figures~\ref{fig11}  and \ref{fig12}). Consequently, CNT (8, 0) nanotube can
be used as toxic gas sensors for CO and NO due to their electronic structure and chemical properties.
This nanotube can be used in part for Cl$_2$ molecule, because contribution to the valence band is very small. For H$_2$S toxic molecule, contribution to the valence band of CNT (8, 0) nanotube is zero. Therefore, pure CNT (8, 0) cannot be used as toxic gas sensors while the B and N doped CNT (8, 0) can be used as toxic gas sensors for H$_2$S molecule.

The geometrical and chemical characteristics of atoms in the CO and NO molecule are very close to the properties of a carbon atom. Hence, toxic gases such as CO and NO can be detected by the CNT (8, 0) nanotube. Atoms in the Cl$_2$ and H$_2$S molecules are far from the carbon atom. In this regard, nanotubes cannot detect such molecules as Cl$_2$ and H$_2$S or can detect them partially. To overcome these restrictions, we can implement different doping procedures (such as B and N) for nanotubes.

Moreover, we have investigated the optimum adsorption behaviors and an optimum binding distance of the proposed CNT structures since we finalize better the physical mechanism of the operation principle of the CNT gas sensors for toxic chemical gases. In this regard, the adsorption energy ($E_{\mathrm{ad}}$) is defined as the difference between the energy of the CNT + molecule(s) system, minus the sum of the energies of the CNT and of isolated molecule(s) in the same conditions. We have calculated the adsorption energy of the system as a function of the distance between the gas molecule and the surface of the CNT wall. The adsorption energy is calculated using the absorption equation [$E_{\mathrm{ad}} = (E_{\mathrm{cnt}}+E_{\mathrm{gas}})-E_{\mathrm{cnt}\text{-}\mathrm{gas}}$]. In this way, we have obtained the adsorption energy and the binding distance of the pure and BN doped CNT (8, 0).

\begin{figure}[!t]
\centerline{\includegraphics[width=0.95\textwidth]{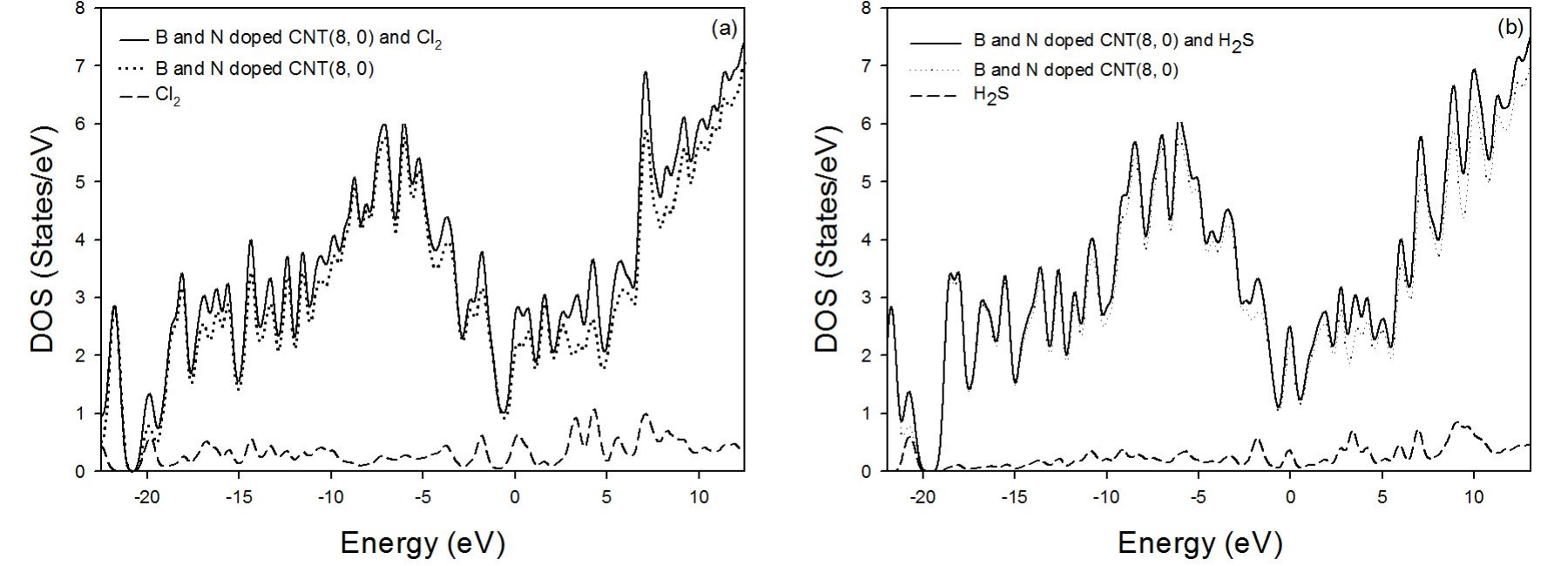}}
\caption{The total and projected density of states for adsorption of (a) Cl$_2$ and (b) H$_2$S on a doped zigzag (8, 0) nanotube. The position of the Femi level is set to zero.} \label{fig13}
\end{figure}

In our calculations we show that the pure CNT exhibits a much enhanced binding with NO and CO, the adsorption energy and the binding distance between CNT and molecules being $-0.394$, $-0.243$~eV and $2.87$, $2.92$~\AA{}, respectively. However, calculations for Cl$_2$ and H$_2$S show a low adsorption energy and binding distance between CNT and molecules which are $-0.095$, $-0.071$~eV and $2.98$, $3.12$~\AA{}, respectively. We can say that a pure CNT (8, 0) is suitable for NO and CO and can be used as a gas sensor. If we examine the doped CNT, our calculations show that the BN-CNT exhibits large adsorption energy values for all proposed toxic chemical gases. The adsorption energy and binding distance are $-6.284$, $-5.591$, $-0.131$, $-0.121$~eV and $1.63$, $1.71$, $1.85$, $2.23$~\AA{} for NO, CO, H$_2$S and Cl$_2$ gases, respectively. Compared to a pure CNT, BN-CNT shows a higher adsorption energy with respect to all the gas molecules considered. These results are in good agreement with the graphics of total and partial densities of states and show that the the BN-CNT can be used as a gas sensor for the proposed gases.

\section{Conclusions}

In the present work, we have made a detailed investigation of structural and electronic properties of pure and doped SWCNTs consisting of zigzag (8, 0), zigzag (10, 0) and zigzag (16, 0) nanotubes using the density functional methods. The results of structural optimization implemented using the LDA are in good agreement with the other theoretical results. At the first stage, we calculated pure and doped nanotubes and investigated the doping effects on carbon nanotubes. At the next step, we found out that the total and partial densities of states of the SWCNT (8, 0) nanotube altered considerably after detecting the CO, NO, Cl$_2$ and H$_2$S toxic molecules. Since we can clearly see the effects of doping, the doping process was performed in two stages: different carbon nanotubes for a constant doping rate and for different doping rates for zigzag (8, 0) carbon nanotubes. Since carbon nanotubes have been recommended for several potential applications such as molecular sensors and nano-electronic devices, we think that the results on the doping rate and type of nanotubes can contribute to subsequent theoretical and experimental studies. Later on, we chose a zigzag (8, 0) nanotube for toxic molecules and investigated the interaction of these toxic molecules with a zigzag (8, 0) nanotube.  There is a transfer of charge from carbon to toxic molecules according to the calculation results, but this condition is valid for the atoms close to a carbon atom. Hence, we can say that  nanotubes can be used as toxic gas sensors for CO and NO molecules.
However, nanotubes cannot detect H$_2$S and can detect Cl$_2$ toxic molecules partially. Therefore, we created the B and N doped CNTs and obtained the total and partial densities of states of these structures. According to these results, B and N doped CNT (8, 0) can be used as toxic gas sensors for such molecules as Cl$_2$ and H$_2$S. In the final stage, we investigated the adsorption behaviors for all toxic molecules. We show that the adsorption results are in good agreement with the graphics of total and partial densities of states. Since there are no experimental data available for the doping rate and toxic gas sensors, we think that an ab initio theoretical estimation is the only reasonable tool to obtain such important data. Consequently, the proposed doping rate methodology and toxic gas sensors can be used in experimental studies and for the purpose of fabricating gas sensors.


\ukrainianpart

\title
{Електронні властивості легованих одностінкових вуглецевих нанотрубок  та датчики  на вуглецевих нанотрубках}
\author{Е. Тетік}
\address{
Університет Мустафи Кемаля, Центр досліджень і застосування комп'ютерних наук,\\
31100 Хатай, Туреччина
}

\makeukrtitle

\begin{abstract}
\tolerance=3000%
Представлено  ab initio обчислення  зонної структури та густини станів напівпровідникових вуглецевих  нанотрубок з однією
стінкою, що володіють високими ступенями  (аж до 25\%) заміщення B, Si і N.
Процес легування складається з двох етапів, а саме, різні вуглецеві нанотрубки для сталої швидкості легування  та різні швидкості легування
для зигзагоподібної  (8, 0) вуглецевої нанотрубки. Проаналізовано  залежність легування  нанотрубок від швидкості легування і від типу
нанотрубки.  На основі цих результатів вибрано  зиг\-загоподібну  (8, 0) вуглецеву нанотрубку  для обчислення датчика токсичного
газу та отримано повні і парціальні густини станів вуглецевих нанотрубок (8, 0).  Показано, що вуглецева нанотрубка (8, 0)
може бути використана в якості  датчиків токсичних газів для молекул  CO і NO; вона здатна частково виявляти токсичні молекули
Cl$_2$, але не здатна виявляти  H$_2$S. Щоб подолати ці обмеження, створено B і N леговану вуглецеву нанотрубку (8, 0)
та отримано повну та парціальну густини станів цих структур. Показано, що  B і N леговані вуглецеві  нанотрубки можуть бути використані
в якості
датчиків токсичних газів для таких молекул як  CO, NO, Cl$_2$ і H$_2$S.
\keywords ab initio обчислення, структура вуглецевих нанотрубок, датчики газу, ефекти легування та заміщення
\end{abstract}

\end{document}